# Nonlinear optics at all-dielectric nanoantennas and metasurfaces

**Basudeb Sain, Cedrik Meier and Thomas Zentgraf**[*]
University of Paderborn, Department of Physics, Warburger Str. 100, 33098 Paderborn, Germany

**Abstract**: Freed from phase-matching constraints, plasmonic metasurfaces have contributed significantly to the control of the optical nonlinearity and enhancing the nonlinear generation efficiency by engineering subwavelength meta-atoms. However, the high dissipative losses and the inevitable thermal heating limit their applicability in nonlinear nanophotonics. All-dielectric metasurfaces, supporting both electric and magnetic Mie-type resonances in their nanostructures, have appeared as a promising alternative to nonlinear plasmonics. High-index dielectric nanostructures, allowing additional magnetic resonances, can induce magnetic nonlinear effects, which along with electric nonlinearities increase the nonlinear conversion efficiency. In addition, low dissipative losses and high damage thresholds provide an extra degree of freedom for operating at high pump intensities, resulting in a considerable enhancement of the nonlinear processes. In this review, we discuss the current state-of-the-art in the intensely developing area of all-dielectric nonlinear nanostructures and metasurfaces, including the role of Mie modes, Fano resonances and anapole moments for harmonic generation, wave mixing, and ultrafast optical switching. Furthermore, we review the recent progress in the nonlinear phase and wavefront control using all-dielectric metasurfaces. We discuss techniques to realize all-dielectric metasurfaces for multifunctional applications and generation of second-order nonlinear processes from CMOS compatible materials.

[*]E-mail: thomas.zentgraf@uni-paderborn.de

## 1 Introduction

Nonlinear effects in electricity and magnetism have been recognized since Maxwell's time. However, much progress has been made in the field of nonlinear optics since the discovery of the laser[1], which made high-intensity optical fields easily feasible. The field started to grow with the first experimental work of Franken and co-workers[2] on optical second harmonic generation in 1961, and the theoretical work of Bloembergen and co-workers[3,4] on optical wave mixing in 1962. Over the following decades, the field of nonlinear optics has witnessed an enormous growth, leading to the observation of new physical phenomena and giving rise to novel concepts and applications including high-harmonics generation and frequency mixing that can act as new light sources or as amplification schemes, light modulators for controlling the phase or amplitude of a light beam, optical switches, optical logic, optical limiters and numerous ways of processing the information content of data images, which created a revolutionary change in photonics technology in the 20$^{th}$ century.[5,6] Almost all those achievements were made on conventional bulk crystals where cumbersome phase-matching conditions limit the efficiency of the nonlinear processes. The current research trend in nonlinear optics has been moved towards miniaturized optical materials in truly compact setups. In recent years, significant advancements in nanofabrication techniques have considerably broadened the experimental and theoretical framework in which the nonlinear optical processes are explored. Major works have been done in design and fabrication to simultaneously address the efficiency and phase matching in nonlinear generation within the sub-wavelength regime, over the last decade. Metamaterials and their two-dimensional counterparts, metasurfaces[7–16], are of great promise for an efficient nonlinear generation of new waves. Metasurfaces can exhibit strong nonlinear optical responses compared to its three-dimensional structures, because of the relaxation or completely overcoming the phase-matching requirement. Freed from phase-matching limitations, and featuring a unique control over nonlinear fields, plasmonic metasurfaces were employed to the fullest extent for the generation of high-harmonics, frequency mixing, and other nonlinear effects.[10,11,16–22] In case of nonlinear plasmonics, the efficiency of the nonlinear optical processes is determined not only by the quality of the phase matching between the interacting optical beams but also the degree of confinement and overlap between the optical near-field and nonlinear optical structures with subwavelength features.[11,18,23–25] Plasmonic materials are most commonly made of metals at the nanoscale. Metal nanostructures (nanoantenna) are variously shaped objects, with size as small as few tens of nanometers, typically made of noble metals, such as gold and silver. Nanostructures supporting surface plasmon polariton resonances that provide both electric field enhancement and spatial confinement, enabling the generation of pronounced nonlinear optical effects at relatively low excitation powers even though the interaction volume might be very small. Plasmonic metasurfaces allow a large degree of control of the optical nonlinearity by engineering subwavelength meta-atoms, enhancing the nonlinear generation efficiency, have been observed during the last decade.[16–18,26–34] However, second-order nonlinear processes, such as second harmonic generation, cannot be achieved from the metasurface having centrosymmetric modes at both the fundamental and generated frequencies. The second-order processes from metal nanostructures originate from two contributions, the bulk and the surface. In noble metal, the local-bulk source is absent because of the material's centrosymmetry, and only the nonlocal-bulk contribution needs to be considered.[35] On the contrary, the local-surface contribution is allowed due to the symmetry breaking at the interface with the embedding medium.[36,37] The magnitudes of the nonlocal-bulk and surface contributions depend on the shape of the nanoparticle and on the optical properties of the metal at the fundamental and second-harmonic frequencies.[38–41] Metasurfaces enabling second harmonic generation (SHG) have been constructed by choosing specific geometries of the meta-atoms such as L-[42–47] and G-shaped antennas[48–52], split-ring resonators[27,53–58], asymmetric dimers[59,60], dielectric-loaded plasmonic 3D structures[61], and multi-resonant antennas[62–65], where the inversion symmetry is absent. Plasmonic metasurfaces have been employed for other second-order processes like sum frequency generation (SFG)[66,67] and difference frequency generation (DFG)[68–70]. In contrast, third-order nonlinear effects, such as the Kerr effect[71–73], third harmonic generation (THG)[29,31,74–77], and the four-wave mixing (FWM)[23,25,78–83], are free from symmetry consideration for linear polarized light. In addition, high-harmonic generation[84] and supercontinuum white light generation[32] have also been realized using plasmonic nanostructures.

Metasurface can locally control the phase, amplitude or the polarization state of light waves that are propagating through or reflected from them. The concept of phase tailoring plasmonic metasurfaces at nonlinear regime enables both the coherent generation and manipulation such as beam steering and lensing of light beams. Nonlinear phase control has been demonstrated for SHG, THG, and FWM in metallic thin films.[24,25,85] Recently, a plasmonic metasurface hologram has been realized at the THG frequency.[31] In addition, nonlinear holography has been demonstrated to be operated at both fundamental and second harmonic frequencies using a Pancharatnam–Berry (PB) phase change, which operates in both the linear and nonlinear optical regimes simultaneously.[27]

So far, we have seen that surface plasmon polaritons are capable to enhance and spatially confine optical fields beyond the diffraction limit. Plasmonic effects in metallic nanostructures have been extensively used to enhance and control the nonlinear optical processes at the nanoscale, such as harmonic generation, wave mixing, supercontinuum generation, nonlinear imaging, and holography, etc. However, several disadvantages limit their applicability in nonlinear nanophotonic applications, including high dissipative losses and inevitable thermal heating, leading to low optical damage thresholds. Thus, the use of all-dielectric metasurfaces supporting magnetic resonances, and the ability to withstand much higher pump field intensities, would be a promising route to obtain higher nonlinear conversion efficiencies.[86] Furthermore, it has been discovered that highly efficient and flexible light manipulation can be achieved at the nanoscale by tuning the electric and magnetic response of all-dielectric nanostructures.[16,87–89] The electric field confinement in dielectric nanoresonators is not limited to the surface only; the additional volume resonance can be added to make the overall enhancement larger.

In this review, we highlight the recent progress in the field of nonlinear optical processes with all-dielectric nanosystems, from nonlinear frequency generation and phase control to applications. The review is organized as follows. In Section 2, we discuss the existence of different resonant modes inside a dielectric nanostructure. In Section 3 and Section 4, we review nonlinear effects based on third- and second-order optical nonlinearities. Section 5 is focused to give insight about the nonlinear switching. Finally, in Section 6 we provide an outlook and future directions in this field.

There is a large number of publications available on nonlinear optical effects in artificial materials including epsilon-near-zero (ENZ) materials, perovskites, 2D materials and multiple-quantum-wells (MQWs). A detailed overview of these topics is well beyond the scope of this review. For a detailed and complete survey, we refer the reader to a well-known review paper on these topics.[16]

## 2  Multipolar resonances in all-dielectric systems

In this section, we will discuss the different modes that are available in all-dielectric nanostructures and their dependence on geometry, which is responsible for nonlinear field enhancement. The optical response of spherically symmetric scatterers, irrespective of their size and constituting medium, can be analytically predicted by expanding the electromagnetic fields in the multipolar basis. This is commonly known to as the Lorenz-Mie theory.[90] For lossless and nonmagnetic materials, their scattering properties can be fully determined when two parameters are specified: the permittivity ε and a size parameter $s$, which is defined as the proportional ratio between the nanoparticle radius R and the wavelength of light λ, $s=2\pi R/\lambda$.[91] In case of sub-wavelength spherical plasmonic scatterers ($s<1$) only electric type resonances can usually be excited and the magnetic response is negligible as the field inside the sphere vanishes, while high refractive index dielectric scatterers exhibit both magnetic and electric type resonances, known as Mie resonances.[88,90,91] The resonant magnetic dipole moments originated from the coupling of incident light to circular displacement current of the electric field, due to the field penetration and phase retardation inside the particle. The magnetic resonance appears when the wavelength inside the particle becomes comparable to its spatial dimension; $2R \approx \lambda/n$, where n is the refractive index of nanoparticle material, R is the nanoparticle radius, and $\lambda$ is the light's wavelength. Mie type resonant behavior is not just specific to spherical scatterers. Non-spherical scatterers such as nanocubes[92], spheroids[93,94], disks and cylinders[95], rings[96], and many other geometries[97] were also shown to support electric and magnetic Mie resonances. This gives the freedom to design various all-dielectric nanostructures with a desirable range of input wavelengths, to achieve resonant conditions. Figure 1(a) shows the schematic of charge-current distributions of the four major resonant modes in high-index dielectric particles (magnetic dipole, electric dipole, magnetic quadrupole, and electric quadrupole).[98] A positive charge (such as a proton) and a negative charge (such as an electron) form an electric dipole, but they are not assumed to be in motion relative to each other, while a magnetic dipole, generally a tiny magnet of microscopic to subatomic dimensions, equivalent to a flow of electric charge around a loop.[99] Electrons circulating around atomic nuclei, electrons spinning on their axes, and rotating positively charged atomic nuclei all are magnetic dipoles. An elementary electric quadrupole can be represented as two dipoles oriented antiparallel. Both the monopole moment (total charge) and dipole moment for this configuration is zero, but there exists a nonzero quadrupole moment. Likewise, a magnetic quadrupole can be realized by employing two pairs of identical current loops such that the dipole moments of both the loops in each pair are antiparallel, while, the pairs are perpendicular to each other. Such a configuration cancels the dipole moment and gives a quadrupole moment. Figure 1(b) illustrates the spectral position of the corresponding modes for silicon nanodisk with height 660 nm and diameter 660 nm in air (see the scanning electron microscopy image in Figure 1(c)).[100] The resonant behavior of subwavelength high-refractive index structures in the visible and near-infrared region was first experimentally demonstrated while studying the optical response of silicon nanowires.[101,102] Later, it has been shown that silicon (Si) nanospheres with sizes ranging from 100 to 300 nm support strong magnetic and electric dipole resonances in the visible and near-IR spectral range, shown in Figures 1(d), (e) and (f).[87] Mie resonators featuring both electric and magnetic responses are seen as a promising platform capable of leading to a practical realization of the Kerker conditions[103,104] (suppression of the back-scattered field under given conditions) with nonmagnetic materials.[105,106] An experimental verification of this effect in high-refractive-index particles was carried out in the microwave range[107] and subsequently observed in the visible range with silicon[93] and Gallium Arsenide (GaAs) nanoparticles[108], where the Kerker effect was due to the interference between the fields radiated by the induced electric and magnetic dipoles. It has been shown that a generalization of this effect to higher-order multipoles is also possible.[109,110]

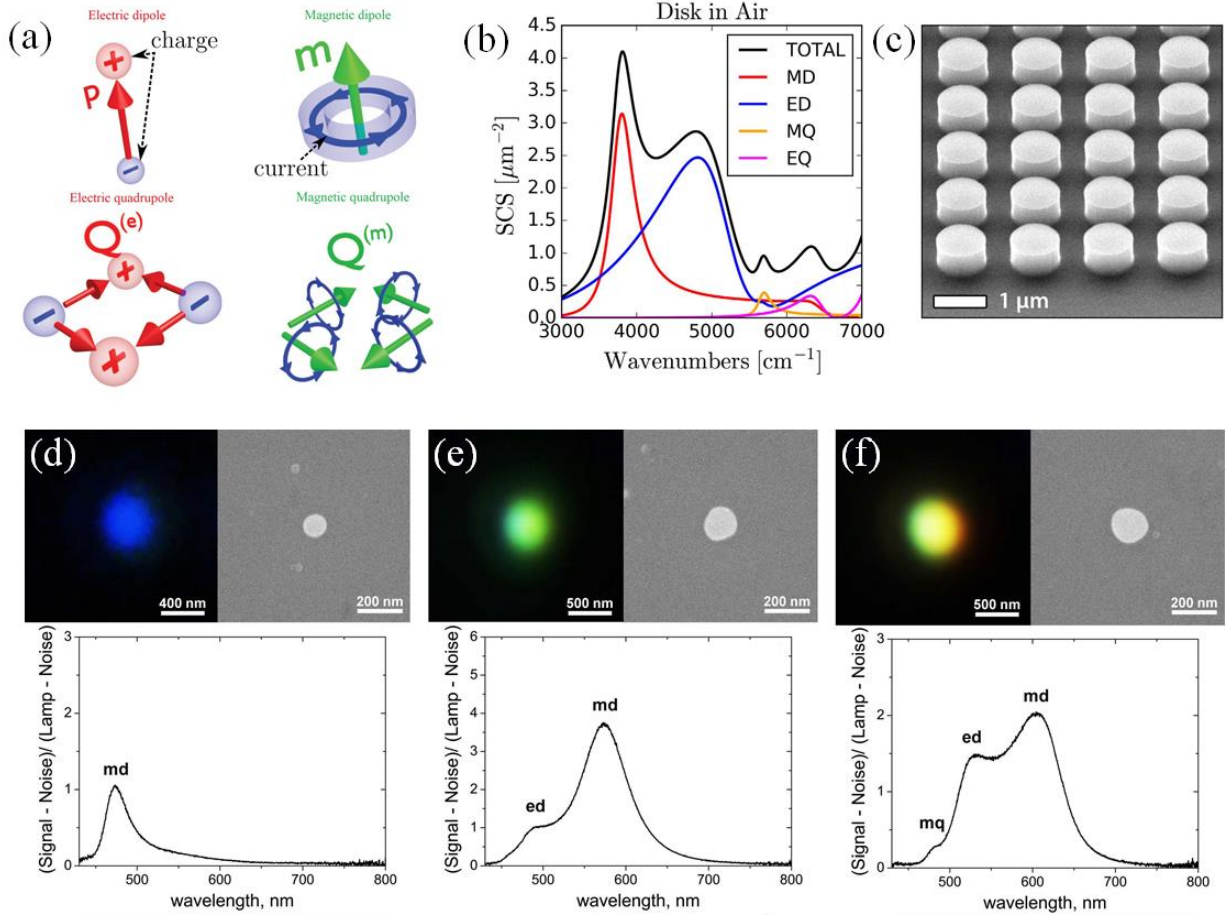

**Figure 1. Mie resonances in dielectric nanostructures. (a)** Schematic illustration of the charge-current distributions that give rise to the electric dipole (p), magnetic dipole (m), electric quadrupole ($Q^{(e)}$) and magnetic quadrupole ($Q^{(m)}$) [Ref. 98]. **(b)** The simulated multipolar decomposition of the scattering-cross-section of an individual an individual Silicon nanodisk with height h = 660 nm and diameter d = 660 nm in air [Ref. 100]. **(c)** Scanning electron microscope (SEM) image of one of the fabricated Silicon disk arrays [Ref. 100]. **(d-f)** Dark-field optical microscope images (top left), SEM images (top right), and dark-field scattering spectra (bottom) of spherical Si nanoparticles with approximate diameters of 100 nm **(d)**, 140 nm **(e)**, and 180 nm **(f)** [Ref. 87]. Figure reprinted with permission: (a) Ref. 98, © 2014 by the American Physical Society (APS); (b) and (c) Ref. 99, © 2016 by the Nature Publishing Group (NPG); (d)–(f) Ref. 87, © 2012 by NPG.

For metallic nanoantennas, the electric dipole modes usually dominate the Mie scattering. In contrast to plasmonics, strong localization of electric and magnetic fields at the nanoscale due to Mie resonances inside dielectric nanoparticles enhances nonlinear effects. It has been acknowledged that the intrinsic microscopic nonlinear electric polarizability of resonant nanoparticles may induce magnetic nonlinear effects.[111] The presence of both electric and magnetic nonlinearities enhances the interference effects, which in turn increases the efficiency and controls the polarization of the nonlinear processes, as well.[89,112]

Another important resonance mode that can be achieved in dielectric nanostructures by possessing more complex design is the Fano resonance.[113,114] The Fano resonance is considered as an asymmetric lineshape of resonances which arises from an interference of discrete (resonance) states with broadband (continuum) states.[113] To observe Fano resonance from all-dielectric nanoparticles/metasurfaces, one of the important concepts is to include interaction between resonant (bright) and nonresonant (dark) scattering modes (Figures 2(a) and (b))[114], which can be recognized as a non-symmetrical dip in the scattering spectrum (Figure 2(c))[114]. Boosting the near field of the resonant nanoparticle at the Fano frequency is considered an important approach to increase nonlinear light-matter interaction. In addition to a strong local field enhancement, the Fano resonance allows controlling the radiative damping of the resonant modes. Besides the electric type of the Fano resonance, all-dielectric nanostructures exhibit a similar magnetic one, related to the optically induced magnetic dipole mode of the individual high-index nanoparticles. This is an additional degree of freedom to manipulate the magnetic resonances of dielectric nanostructures to enhance the nonlinear interaction.

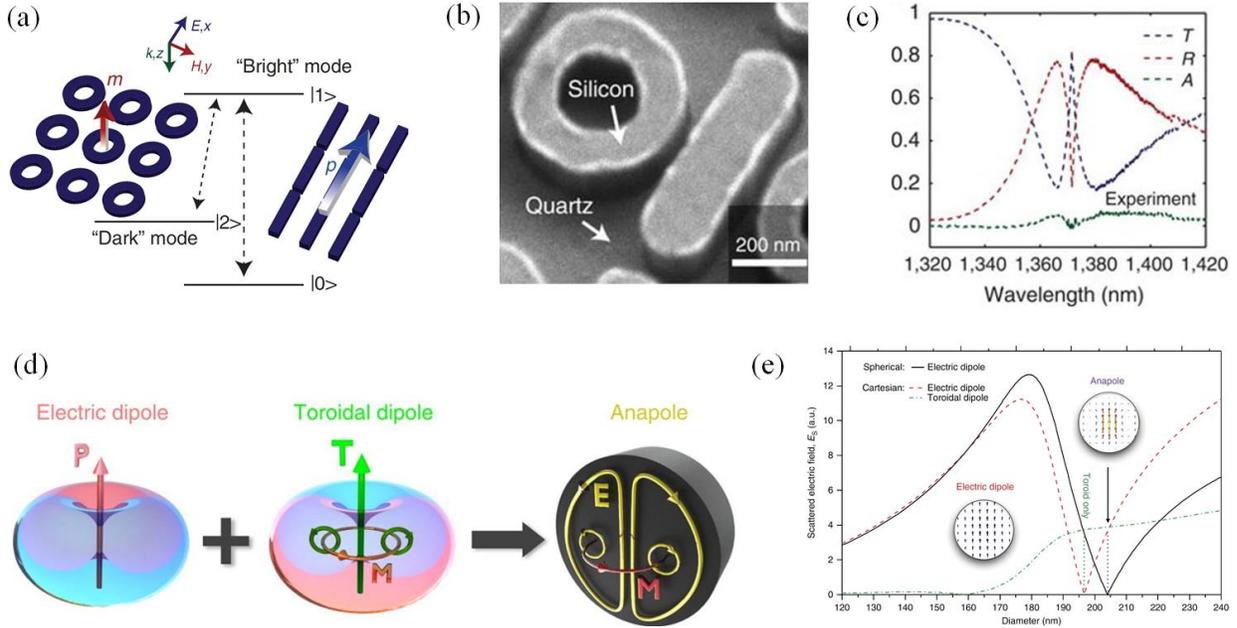

**Figure 2. Fano resonances and anapole modes in dielectric nanostructures.** (**a**) Schematic illustration of the interference between the bright- and dark-mode resonators, (**b**) corresponding SEM image of a single unit cell of the fabricated metasurface and (**c**) corresponding experimental transmittance (**T**), reflectance (**R**) and absorption (**A**) spectra, showing a Fano-type resonance [Ref. 114]. (**d**) Schematic illustration of an anapole excitation: The toroidal dipole moment is associated with the circulating magnetic field M accompanied by electric poloidal current distribution. Since the symmetry of the radiation patterns of the electric P and toroidal T dipole modes are similar, they can destructively interfere leading to total scattering cancelation in the far-field with non-zero near-field excitation [Ref. 115]. (**e**) calculated spherical electric dipole (black), Cartesian electric (red) and toroidal (green) dipole moments contributions to the scattering by a dielectric spherical particle of refractive index n=4 and wavelength 550 nm, as a function of the diameter. The anapole excitation is associated with the vanishing of the spherical electric dipole when the Cartesian electric and toroidal dipoles cancel each other [Ref. 115]. Figure reprinted with permission: (a)–(c) Ref. 114, © 2014 by NPG; (d) and (e) Ref. 115, © 2015 by NPG.

High-index dielectric nanoparticles also support other unusual electromagnetic scattering modes such as anapole modes.[115–117] Anapoles are characterized by a specific configuration of excited fields inside a system. When the toroidal and electric dipole modes spectrally overlap, they produce almost equivalent radiation patterns in the far field but with opposite phases, generating a pronounced dip in the spectrum (Figure 1(e))[115], with non-vanishing near-field.[115,117–119] The lack of scattering and radiation loss in a dipole channel can further enhance the local fields, boosting nonlinear effects. The recent development of all-dielectric nonlinear nanostructures that can show comparable electric and magnetic multipolar contributions have led to advances in the emerging field of multipolar nonlinear nanophotonics.

## 3   Third-order nonlinear all-dielectric nanostructures and metasurfaces

A wide range of theoretical and experimental studies of nonlinear plasmonics have already laid the foundation of modern nonlinear optics with nanostructures. However, all-dielectric arrangements can support even stronger nonlinear optical responses as well as novel functionalities enabled by signified magnetic dipole and higher-order Mie-type resonances, compared to their plasmonic counterparts. In this section, we present an overview of state-of-the-art progress in the area of nonlinear interactions of high-index dielectric nanostructures and metasurfaces, supporting additional magnetic resonances. In addition, dielectric nanostructures are able to withstand much higher pump fields making them promising to obtain higher nonlinear conversion efficiencies.[89,91,120] The electric field enhancement in dielectric nanostructures is typically smaller than in the plasmonic ones, however, the additional volume resonance can be added to make the overall enhancement larger, as the field confinement in dielectric nanostructures is not restricted to the surface only as in their metallic counterparts.

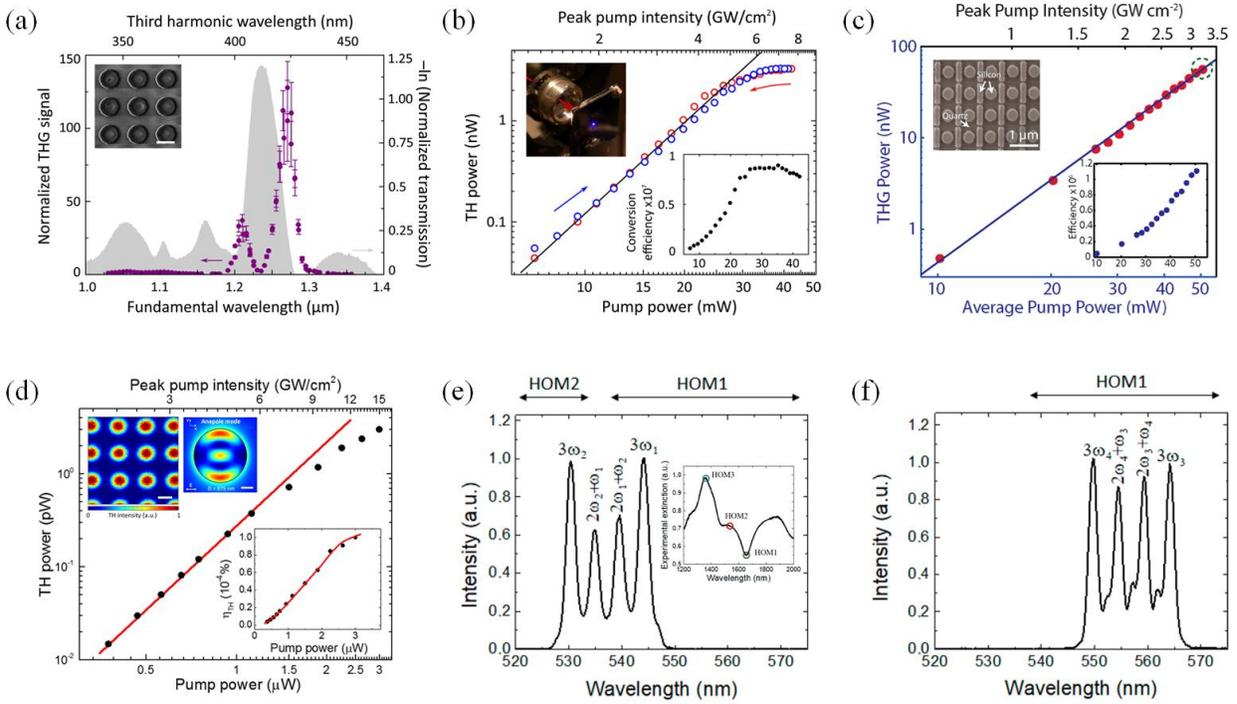

**Figure 3. Third-order nonlinear effects.** (a) THG spectroscopy of Si nanodisk arrays. The negative logarithm of the normalized transmission spectrum of the sample with period p = 0.8 μm, height h = 220 nm, and diameter d = 0.5 μm is shown by the gray area indicating a resonance at 1.24 μm. The THG spectrum of the sample (purple dots) shows a strong enhancement within the spectral band of the resonance. The inset shows the SEM image of the sample [Ref. 121]. (b) Power dependence and conversion efficiency of the resonant THG process in Si nanodisks. Blue circles denote the THG power dependence obtained at λ = 1.26 μm fundamental wavelength. Left inset: photographic image of the sample irradiated with the invisible IR beam. The blue point is the scattered THG signal. Right inset: conversion efficiency of the nanodisk sample as a function of the pump power [Ref. 121]. (c) THG power as a function of the pump power for the Fano-resonance metasurface. The red circles indicate the measured data, and the blue line is a numerical fit to the data with a third-order power function. Left inset: SEM image of the sample; right inset: extracted absolute THG efficiency [Ref. 122]. (d) Measured THG power versus for the excitation of the anapole mode in silicon nanodisks. Left inset: THG intensity image taken at $\lambda_{pump}$ = 1650 nm; scale bar is 1 μm and top view of the simulated distribution of the electric field intensity for a disk diameter of 875 nm. Right inset: conversion efficiency as a function of pump power [Ref. 127]. (e) Measured nonlinear response of a Ge disk when exciting at HOM1 and HOM2 modes simultaneously. Inset: extinction spectrum of a Ge disk of 200 nm height and 625 nm radius [Ref. 129]. (f) Measured nonlinear response of the Ge disk when exciting at two different wavelengths comprising HOM1 [Ref. 129]. Figure reprinted with permission: (a) and (b) Ref. 121, © 2014 by the American Chemical Society (ACS); (c) Ref. 122, © 2015 by ACS; (d) Ref. 127, © 2016 by ACS; (e) and (f) Ref. 129, © 2017 by ACS.

Shcherbakov et al. demonstrated a strong nonlinear response from dielectric nanostructures made of silicon nanodisks.[121] They exhibited enhanced third-harmonic generation (THG), which was observed by naked eye using both isolated and an array of nanodisks that were optically pumped in the vicinity of the magnetic dipole resonance as shown in Figure 3(a) and (b)[121]. The nanodisks were fabricated using a silicon-on-insulator (SOI) wafer and exhibiting both electric and magnetic dipole resonance, for which the silicon metasurface generated up to 4 nW THG power for a pump power of 30 mW (peak pump intensity 5.5 GW cm$^{-2}$). The resulting conversion efficiency of $0.9 \times 10^{-7}$ (Fig. 3(b))[121] was fundamentally limited by free carriers generated via two-photon absorption in the bulk Si substrate that leads to free-carrier absorption of the pump beam.

Third harmonic generation from a Fano nonlinear metasurface consisting of resonant Si nanodisks and nanoslits, supporting resonant dark (magnetic dipole) and bright (electric dipole) modes, respectively, was demonstrated by Yang et al.[122] The nanostructures were fabricated by electron beam lithography (EBL) followed by reactive-ion etching (RIE) after depositing a 120-nm-thick poly-Si layer on a quartz substrate. The measured conversion efficiency was $1.2 \times 10^{-6}$ with an average pump power of 50 mW at a peak pump intensity of 3.2 GW cm$^{-2}$ (Figure 3(c))[122]. The enhanced nonlinearity arises from high-quality factor Fano resonance that in turn strongly enhances the local electric field within the Si, thus resulting in a large effective third-order nonlinearity. Fano resonances can also be excited from nanodisks only by using different lattice arrangement. A square array of symmetric clusters of four Si nanodisks, forming quadrumers, exhibited multifold enhancement of the THG signal, excited by an oblique plane wave.[123] The origin of the Fano resonance in Si nanodisk quadrumers is the destructive interference between the coupled magnetic-like modes formed by out-of-plane magnetic dipoles and circulating displacement current produced by in-plane electric dipoles in the far-field. In addition, the Fano-assisted THG in Si nanodisk trimers was demonstrated.[124] Another example of enhanced THG in a Fano resonant silicon metasurface due to the trapped mode supported by the high quality factor, was demonstrated by Tong et al.[125] The conversion efficiency was enhanced at about 300 times with respect to the bulk silicon slab, which depends on both the wavelength and polarization angle of the pump light.

Benefited from the high damage threshold of all-dielectric nanostructures, a silicon metasurface, created by means of laser-induced self-organized nanostructuring of thin Si films, was employed to generate 30-fold enhanced third-order nonlinear response, demonstrating UV femtosecond laser pulses at a wavelength of 270 nm with a high peak and average power ($10^5$ kW and 1.5 μW, correspondingly).[126]

Germanium (Ge) is another excellent material for nonlinear metasurfaces because of its high refractive index in the visible range and large third-order susceptibility. THG in thin Ge nanodisks under normally incident laser excitation can be boosted via a nonradiative anapole mode (AM). Grinblat et al.[127] demonstrated strong THG by exciting a Ge nanodisk near the anapole mode (Figure 3(d)) and the measured TH intensity was about one order of magnitude larger than that corresponding signal for the excitation of the dipolar resonances, at which the field is poorly confined within the dielectric material. The observed conversion efficiency was ≈$10^{-4}$ upon 1μW (15 GW cm$^{-2}$) pump power. Later the same group demonstrated THG using higher-order anapole modes (HOAM)[128] and four-wave mixing (FWM) using high-order modes (HOMs)[129] that do not show anapole characteristics. In case of FWM, when the two excitation wavelengths were chosen with two different HOMs and the near-field intensity overlap between those modes was about 80% within the disk, the FWM signals were found to be >30% lower in intensity compared to the THG of the individual pump wavelengths (Figure 3(e))[129]. However, when the two different pump wavelengths covering a single HOM, the degenerate FWM signals were observed to decrease by only ~10% in intensity with respect to the THG process, indicating nearly equivalent efficiency (Figure 3(f))[129].

Very recently, Wang et al.[130] demonstrated a new concept for embedding any functionality into a nonlinear all-dielectric metasurface made of silicon, producing phase gradients over a full 0-2π phase range based on the generalized Huygens' principle that was extended to nonlinear optics. Efficient wavefront control of a third-harmonic field, along with the generation of nonlinear beams at a designed angle and the generation of nonlinear focusing vortex beams have been shown in that work (Figure 4)[130].

So far, we have seen that the choice of the appropriate confined optical mode and mode overlap (in the case of wave mixing) are the two utmost important factors to get maximum conversion efficiency. These investigations reveal useful pathways for the further optimization of third-order optical processes in all-dielectric nanostructures.

## 4 Second-order nonlinear all-dielectric nanostructures and metasurfaces

In section 3, we have shown that Si and Ge nanostructures and metasurfaces can be utilized to enhance third-order nonlinearities. However, Si and Ge do not possess bulk-mediated second-order nonlinearities due to their centrosymmetric crystal structure. To overcome this limitation, nanostructures made out of III-V semiconductors that possess a high dielectric index and relatively large second-order susceptibilities have been used.[131]

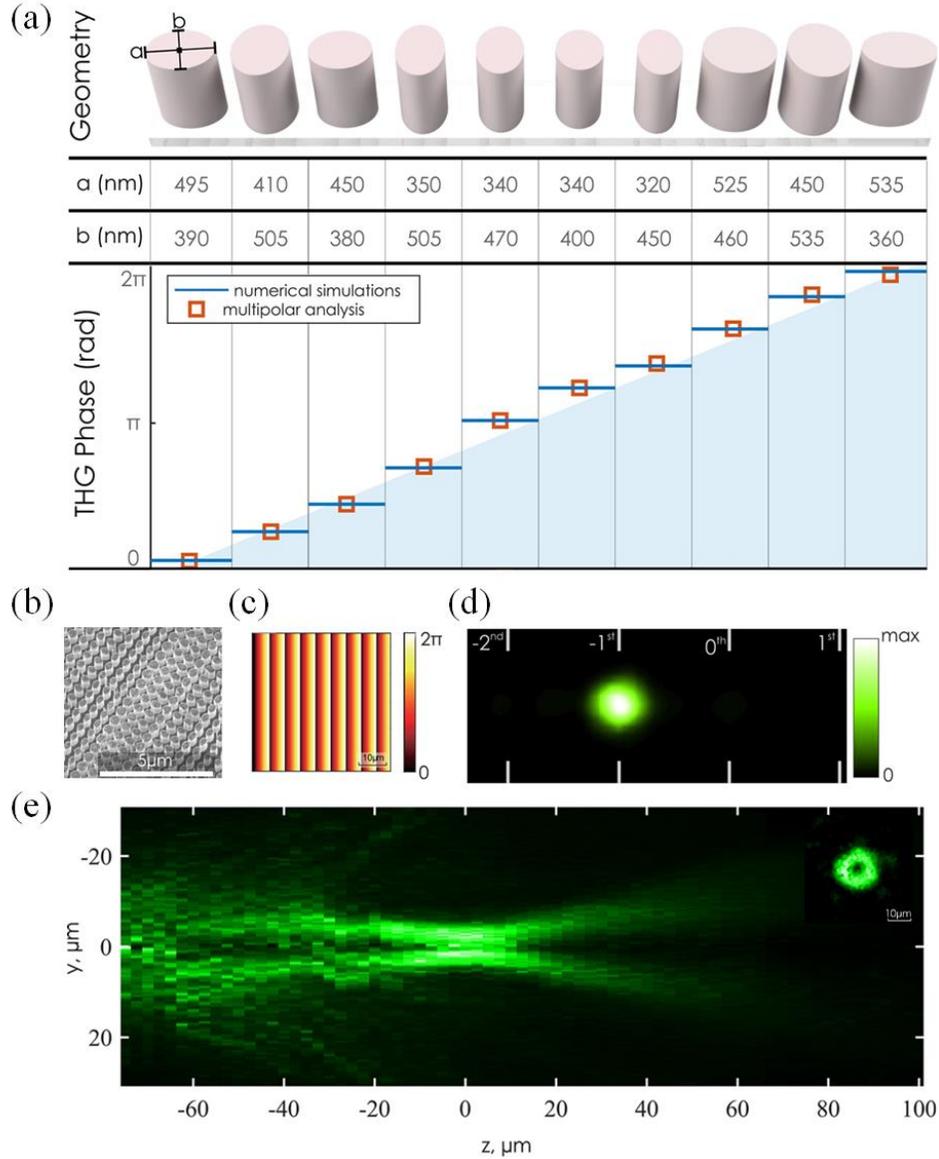

**Figure 4. Nonlinear phase control with silicon metasurfaces.** (a) Geometries and nonlinear phases of Si nanopillar meta-atoms. Shown are the sizes of the nanopillars and corresponding analytical and numerical results for the phase of the third-harmonic field for a pump wavelength of 1615 nm and linear polarization of the pump along the a-axis. (b) SEM image of the silicon metasurface. (c) Phase profile of the THG field encoded into the metasurface. (d) k-space image of the forward THG signal. A total of 92% of TH is directed into the designed diffraction angle θ = 5.6°, where $k_x/k_0$ = -0.098. (e) Cross-section of a generated donut-shape vortex beam at the THG taken along the propagation direction behind the metasurface. Inset: Cross-section perpendicular to the optical axis at distance z = 25 μm [Ref. 130]. Figure reprinted with permission: (a)–(e) Ref. 130, © 2018 by ACS.

Resonantly enhanced SHG using gallium arsenide (GaAs) based dielectric metasurfaces, made of arrays of cylindrical resonators, have been demonstrated SHG enhancement factors as large as $10^4$ compared to unpatterned GaAs.[132] The SHG measurements were performed in reflection geometry because the SHG wavelengths were above the bandgap of GaAs so that the SH signal in the transmission direction would have been completely absorbed by the GaAs substrate. The strongest SHG effect was observed when pumped at the magnetic dipole resonance, at which the absolute nonlinear conversion efficiency reaches $\sim 2 \times 10^{-5}$ with $\sim 3.4$ GW cm$^{-2}$ pump intensity, shown Figure 5(a)[132]. Interestingly, the demonstrated conversion efficiency at the magnetic dipole resonance is about ~100 times higher than the conversion efficiency at the electric dipole resonance, which is caused by increased absorption of GaAs at the shorter wavelength of the electric dipole resonance.

Recently, the same group demonstrated a GaAs metasurface based optical frequency mixer (Figure 5(b)) that concurrently generates eleven new frequencies spanning the ultraviolet to near-infrared.[133] The even and odd higher-order nonlinearities of GaAs enabled the observation of SHG, THG, and fourth-harmonic generation (FHG), sum-frequency generation (SFG), two-photon absorption-induced photoluminescence (TPA-PL), FWM and six-wave mixing (SWM) as shown in Figure 5(c)[133]. The resonantly enhanced frequency mixing was achieved by simultaneously exciting the lowest order magnetic and electric dipole Mie resonances of a GaAs nanocylinder. The simultaneous occurrence of these seven nonlinear processes is assisted by the beams [Ref. 133]. Figure reprinted with permission: (a) Ref. 132, © 2016 by ACS; (b) and (c) Ref. 133, © 2018 by NPG.

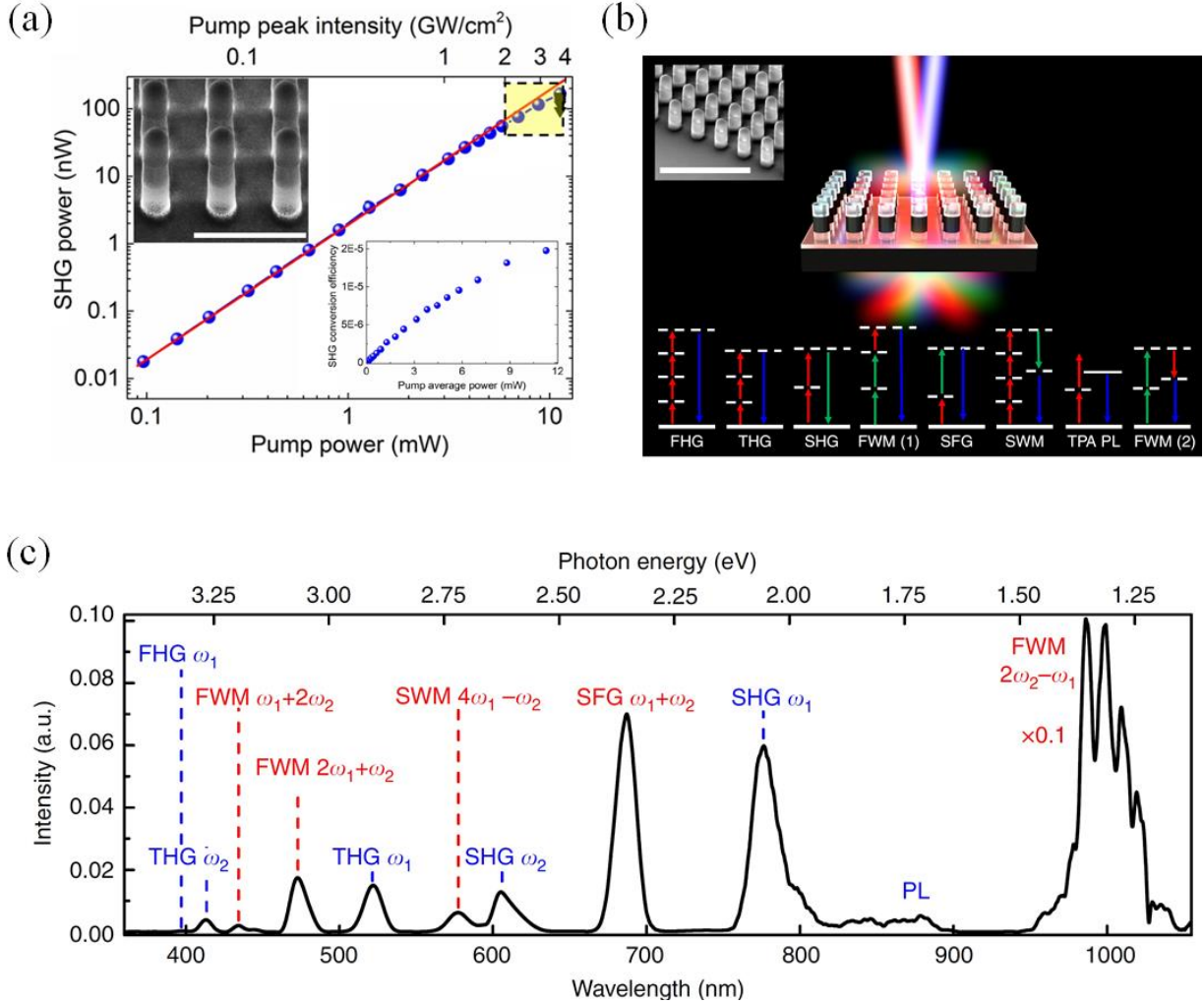

**Figure 5. Second-order nonlinear effects at GaAs metasurfaces.** (**a**) SHG power dependence at low pump intensities, and the deviation from the quadratic relationship at higher pump intensities due to the damage of GaAs resonators. Left inset: SEM images of the fabricated GaAs resonator array. Right inset: SHG conversion efficiency as a function of pump power [Ref. 132]. (**b**) Schematic illustration of an optical metamixer consisting of a square array of subwavelength GaAs dielectric resonators. Two femtosecond near-IR pulses pump the metamixer and a variety of new frequencies are simultaneously generated. Top inset: SEM image of the GaAs metamixer (scale bar 3 μm). Bottom inset: energy diagrams of the seven nonlinear optical processes that occur simultaneously at the metasurface: second-harmonic generation (SHG), third-harmonic generation (THG), fourth-harmonic generation (FHG), sum-frequency generation (SFG), two-photon absorption-induced photoluminescence (TPA PL), four-wave mixing (FWM) and six-wave mixing (SWM) [Ref. 133]. (**c**) Measured nonlinear spectrum exhibiting eleven generated peaks originating from seven different nonlinear processes when two optical beams at $\lambda_2 \sim 1.24$ μm and $\lambda_1 \sim 1.57$ μm are used to simultaneously pump the GaAs metasurface. Blue labels indicate harmonic generation processes and photoluminescence arising from two-photon absorption that each requires only one pump beam. Red labels indicate frequency mixing that involves both pump

combined effects of strong intrinsic material nonlinearities, enhanced electromagnetic fields, and relaxed phase-matching requirements. The results illustrate that metasurfaces can be a versatile system to achieve multiple nonlinear processes with equal contributions simultaneously.

Shaping the unidirectional SHG radiation pattern from aluminum gallium arsenide (AlGaAs) nanodisk antennas as well as its polarization state, generation of cylindrical vector beams of complex polarization have been experimentally demonstrated.[112] In these experiments, nonlinear conversion efficiencies exceeding $10^{-4}$ have been achieved. In another work, SHG from monolithic AlGaAs optical nanoantennas of optimized geometry, excited by a magnetic dipole resonance at the wavelength of 1550 nm, has been measured, revealing a peak conversion efficiency exceeding $10^{-5}$ at 1.6 GW cm$^{-2}$ pump intensity.[134]

In an unconventional way, Bar-David et al. reported the generation of SH signal from an amorphous silicon metasurface by their very recent publication.[135] The SH signal was generated mostly from the surface, following the selection rules that rely on the asymmetry of the meta-atoms.

The superiority of the fabricated materials is utmost important to get efficient nonlinear phenomena. Fabrication of the dielectric metasurfaces of nonzero second-order bulk susceptibility requires special attention to maintain their high quality, as they made of III-V semiconductor nanostructures. In this context, widegap materials, such as ZnO, GaN or LiNbO$_3$, allowing even lower losses at shorter wavelengths and can be an alternative as second-order materials to realize highly nonlinear all-dielectric metasurfaces.

The high-index dielectric metasurfaces provide a strong nonlinear response, low dissipative losses and high damage threshold. These advantages make them a powerful platform for modern nonlinear nanophotonics. The presence of both the electric and magnetic responses makes it possible to tune the scattering patterns and design switchable flat optical devices engaging these nonlinearities.

## 5   All-dielectric ultrafast optical switching

One of the biggest advantages of metasurfaces is the ability to spatially vary and tune the optical parameters of the surface. Such spatial variations enable new opportunities for the observed ultrafast optical switching, namely to construct ultrafast displays that can switch between two or more different images at the femtosecond timescale. Ultrafast optical switching that is based on the free carrier nonlinearity in semiconductors suffers from long switching time (limited to tens of picoseconds) due to two-photon absorption and comparatively large free carrier lifetime.[136–139] In the past decade, plasmonic metasurfaces provided important progress on optical ultrafast switching based on strong light localization within subwavelength mode volume, which in turn increased the third-order nonlinearity, resulting in a change of the complex refractive index of the material.[140–152] However, optical loss and heating effects in plasmonic nanoantennas limit the device performance. In this context, high-permittivity all-dielectric metasurfaces can be a promising alternative. In this section, we discuss the recent progress of ultrafast switching effect using all-dielectric metasurface. Makarov et al.[153] presented an approach for efficient tuning of optical properties of a high refractive index subwavelength nanoparticle. The nanoparticles showed a magnetic Mie-type resonance that was shifted by femtosecond laser irradiation. The effect is based on ultrafast photoinjection of a dense ($>10^{20}$ $cm^{-3}$) electron-hole plasma within the nanoparticle, drastically changing its transient dielectric permittivity. The work demonstrated experimentally 20% switching of reflection of a single silicon nanoparticle photoexcited by femtosecond laser pulses with a wavelength in the vicinity of the magnetic dipole resonance, enabling high-efficient light manipulation on the subwavelength scale (Figure 6(a))[153]. Later, the same group reported on the experimental observation of a ∼2.5 ps operation regime of a nonlinear all-dielectric nanoantenna, which was an order of magnitude faster than their previous work.[154] A corresponding theoretical study on silicon nanoparticle dimers for nonlinear optical tuning, enabled by photoexcitation of electron-hole plasma, has been studied in another work.[155]

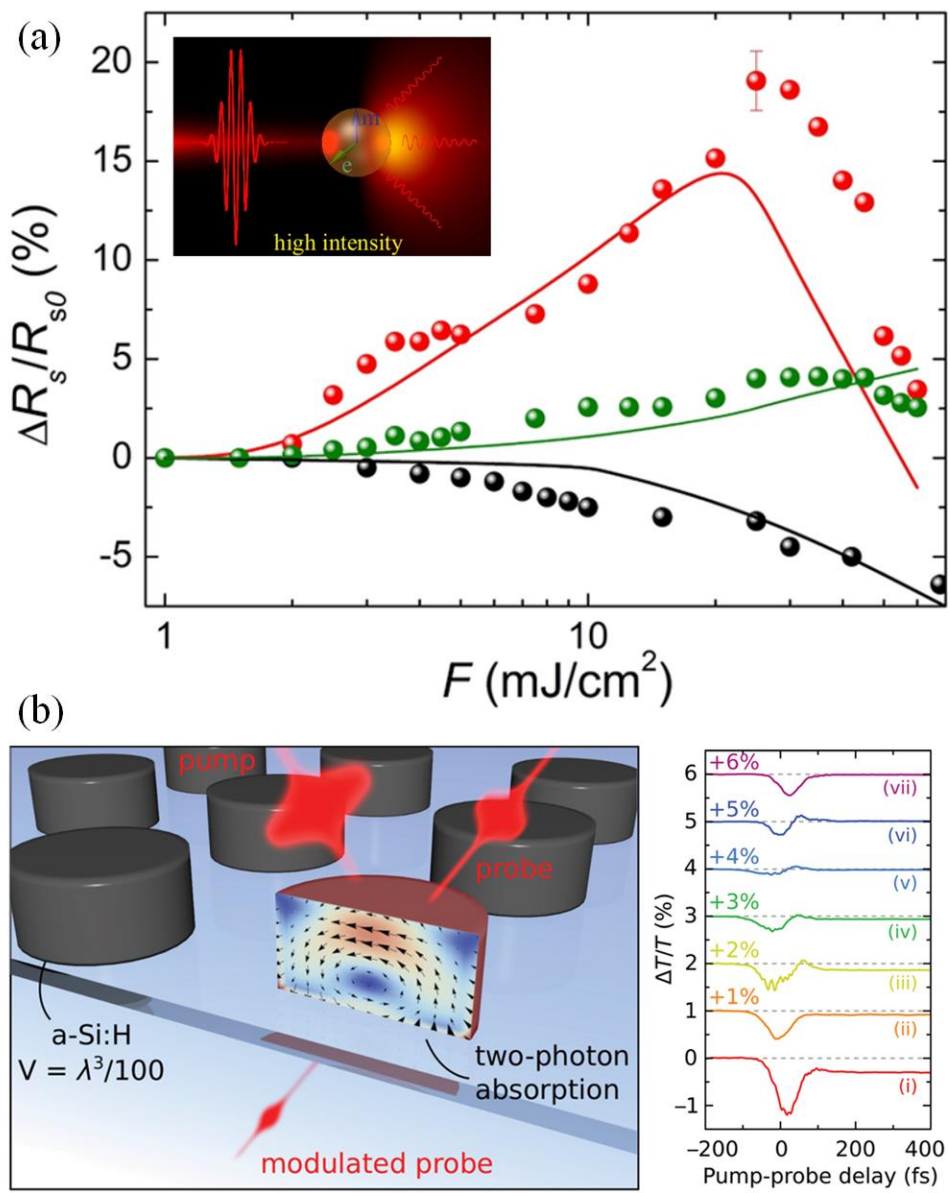

**Figure 6. Ultrafast optical switching with silicon metasurfaces.** (a) Experimental (dots) and theoretical (solid lines) dependencies of the normalized reflectance change on the laser fluence for a 220-nm-thick silicon film (black), the "near-resonance nanoparticle" (red) and the "off-resonance nanoparticle" (green). Inset: schematic illustration of the scattering manipulation by an intense femtosecond laser pulse. The intense laser pulse switches the scattering of the particle to a Huygens source regime when the incident light is scattered in the forward direction [Ref. 153]. (b) Left: Illustration of the ultrafast all-optical switching in resonant silicon nanodisks based on two-photon absorption. Right: Tailoring the all-optical switching in silicon nanodisks. Shown are the relative transmission changes for different samples [Ref. 156]. Figure reprinted with permission: (a) Ref. 154, © 2015 by ACS; (b) Ref. 155, © 2015 by ACS.

All-optical switching of femtosecond laser pulses passing through subwavelength silicon nanodisks at their magnetic dipolar resonance was presented.[156] Pump-probe measurements revealed that the switching of the nanodisks can be governed by bandwidth-limited 65 fs-long two-photon absorption. The authors observed an improvement of the switching time by a factor of 80 with respect to the unstructured silicon film (Figure 6(b))[156]. The undesirable free-carrier effects can be suppressed by proper spectral positioning of the magnetic resonance, making such a structure the fastest all-optical switch operating at the nanoscale.

All-dielectric metasurfaces, benefited from very low intrinsic losses and localized Mie-type modes, are promising for all-optical switching and modulation. Magnetic resonances in all-dielectric metasurfaces suppress the free carrier effect, leading to greatly reduced all-optical switching times without suffering from a strong loss in modulation depth.

## 6  Summary and outlook

In this paper, we have reviewed the state-of-the-art in the intensely developing area of all-dielectric nonlinear nanostructures and metasurfaces, as a promising alternative of nonlinear plasmonic metasurfaces. We have discussed the important role of the electric and magnetic dipole and higher-order Mie modes, in harmonic generation, wave mixing, ultrafast optical switching, including Fano resonances and anapole moments. Electric and magnetic resonances and their interference in high-index dielectric nanostructures strongly influence the enhancement of the nonlinear optical interactions. Although the electric field enhancement in dielectric nanostructures is smaller than in the plasmonic counterparts, the additional volume resonance, coming from the field confinement of the mode in the high-index resonators, can make the overall enhancement of the nonlinear process larger. High-index dielectric nanostructures and metasurfaces, supporting additional magnetic resonances, can induce magnetic nonlinear effects, which along with electric nonlinearities increase the nonlinear conversion efficiency.

Additionally, low dissipative losses and high damage threshold of all-dielectric nanosystems provide an additional degree of freedom in operating at high pump intensities, resulting in considerable enhancement of the nonlinear processes. In comparison to plasmonic nanostructures, this is a huge advantage as the loss and thermal heating effects are mostly undesired and can lead for metallic structures easily to the destruction of the nanostructures.

Despite the tremendous progress in the enhancement of the nonlinear efficiency, much less advancement has been achieved in realizing functional nonlinear all-dielectric metasurface elements. Very few examples are available in the literature about nonlinear phase and wavefront control to show novel optical functionalities. The work by Wang et al. shows that a wavefront control of the third-harmonic field based on the generalized Huygens' principle (which is extended to nonlinear optics) seems feasible.[130] Using such Huygens' Principle for nonlinear processes while keeping the nonlinear conversion efficiency high seems to be an important research aspect for future improvements. Furthermore, the spatial control of the nonlinear phase of the THG signals depends sensitively on the precise geometry and refractive index of the nanostructures, resulting in challenging fabrication. Here, different concepts for the control of the nonlinear phase might bring further advantages. In this context, an elegant way to arbitrarily tailor the nonlinear phase would be based on the Pancharatnam–Berry (PB) phase technique, which was demonstrated for nonlinear effects at plasmonic metasurfaces.[11,24] The PB phase manifests as an accumulated phase during the change of the polarization state for light, for example, if light with a particular polarization is scattered at a nanostructure. Because these phases depend solely on the elements' orientation, it can be interpreted as being of geometrical nature and it is often referred to as a geometrical phase. The concept was previously applied to encode phase information into planar flat surfaces with plasmonic nanostructures, giving rise to nonlinear optical holography, image generation, and beam profile manipulation.[11] We note that the same symmetry selection rules for nonlinear processes as for plasmonic nanostructures are valid, resulting in symmetry dependent nonlinear processes. By tailoring the rotation angle of each nanostructure, the angle will determine the local phase for the nonlinear material polarization. Hence, using the control over the nonlinear PB phase the local phase in the generation process can be controlled. This way, one can generate different nonlinear functional elements that rely on a space-dependent phase of the generated nonlinear signal. One important application of tailoring the nonlinear phase is nonlinear holography. In this context, two or more nonlinear processes can simultaneously be overlapped to create nonlinear holographic multiplexing with different frequencies.

Apart from the conventional selection of the second-order nonlinear materials, the fabrication of metasurfaces is rather complex, another promising direction is to use complementary metal–oxide–semiconductor (CMOS)-compatible materials (such as Si, SiN, $SiO_2$, and Ge) to realize second-order processes by breaking their local symmetry. Second-order nonlinear metasurfaces are required for important processes such as phase-only modulation, sum and difference frequency generation, besides SHG. Other than that they might find important application in quantum nonlinear optics. The symmetry breaking might be possible by applying an external direct current (d.c.) field, similar to electric-field-induced second harmonic generation (EFISHG).[157–159] In this process, the third-order nonlinear susceptibility $\chi^{(3)}$ is converted to a second-order $\chi^{(2)}$ which can introduce a phase shift known as d.c. Kerr effect, an inherently phase-matched process.[160] Recently, Timurdogan et al. demonstrated EFISHG along with the d.c. Kerr effect in integrated silicon ridge waveguides by breaking the crystalline symmetry of silicon through applying d.c. fields and inducing a $\chi^{(2)}$ which is proportional to the $\chi^{(3)}$ of silicon.[161] The $\chi^{(2)}$ originated from the large $\chi^{(3)}$ of silicon combined with large electric fields generated within reverse-biased p–i–n junctions. To achieve an efficient EFISHG in silicon, the fundamental pump and signal modes were quasi-phase-matched with periodically patterned p–i–n junctions.

All-dielectric metasurfaces have a high potential for enabling the efficient generation of new frequencies by using simultaneously more than one nonlinear processes. In such a way, one can construct holographic multiplexing elements based on frequency or polarization. Nonlinear all-dielectric nanosystems might also drive rapid progress in engineering nonlinear optical effects beyond the diffraction limit and has enormous potential to develop new concepts of miniaturized efficient nonlinear photonic metadevices in recent future.

**Acknowledgments**

This project has received funding from the European Research Council (ERC) under the European Union's Horizon 2020 research and innovation programme (grant agreement No. 724306) and the Deutsche Forschungsgemeinschaft (DFG, German Research Foundation) (No. 231447078-TRR142).